\pdfoutput=1
\documentclass[journal]{IEEEtran}

\usepackage{amssymb}
\usepackage{cite}
\usepackage{graphicx}
\usepackage{epsfig}
\usepackage{epstopdf}
\usepackage{algpseudocode}
\usepackage[linesnumbered,ruled,vlined]{algorithm2e}
\usepackage{amsmath}
\usepackage{relsize}
\usepackage{mathtools}
\usepackage{multirow}
\usepackage{pifont}
\usepackage{enumerate}
\usepackage{setspace}
\usepackage{amsthm}
\usepackage{mathrsfs}
\usepackage{array}
\usepackage{xcolor}
\usepackage{caption}
\usepackage{subcaption}
\usepackage{float}
\usepackage[inline]{enumitem}
\usepackage{url}
\usepackage{amsmath,amsfonts,amssymb,amsthm, bm}
\graphicspath{{Images/}{Results/}}

\usepackage{array}
\newcolumntype{L}[1]{>{\raggedright\let\newline\\\arraybackslash\hspace{0pt}}m{#1}}
\newcolumntype{C}[1]{>{\centering\let\newline\\\arraybackslash\hspace{0pt}}m{#1}}
\newcolumntype{R}[1]{>{\raggedleft\let\newline\\\arraybackslash\hspace{0pt}}m{#1}}

\theoremstyle{plain}

\theoremstyle{remark}

\usepackage{mathtools}
\usepackage{xcolor}
\usepackage{array}

\begin{document}
	
	\title{Enabling Efficient Client Selection in FL-as-a-Service for Multi-Application based Society 5.0}
    
	\author{ \begin{tabular}{cc} Prachi Nandi & Sonakshi Satpathy \\ \texttt{prachinandi237@gmail.com} & \texttt{sonakshi1901@gmail.com} \\[1em] Timam Ghosh & Arijit Roy \\ \texttt{timam.ghosh@yahoo.com} & \texttt{arijitroy@iitp.ac.in} \end{tabular} }\maketitle
	
	\begin{abstract}
         The rapid development of the Internet of Things (IoT) has led to the generation of vast amounts of data from sensors, prompting the need for advanced learning models to analyze this data for personalized services. Federated learning (FL) emerges as a solution, offering decentralized learning that preserves user privacy by building models on local devices and sharing only aggregated insights. This paper explores FL-as-a-service (FLaaS) in IoT, highlighting its potential for collaborative learning across applications while addressing challenges like security, privacy, and optimizing hierarchical architectures for efficient model convergence and accuracy. To enhance the effectiveness of FL in IoT environments, this study focuses on selecting optimal nodes for model processing through the evaluation of parameters such as delay, energy consumption, and link status. The proposed method aims to identify suitable client nodes for model training. Performance evaluation is conducted using a Human Activity Recognition dataset under simulated IoT network conditions. The proposed approach is compared against randomized and Q-learning based client selection strategies. Experimental results shows improvements in delay and energy efficiency while maintaining communication performance. The findings highlight the potential of efficient client selection mechanisms for enhancing FLaaS in dynamic IoT ecosystems and supporting future intelligent services in Society 5.0.
	\end{abstract} 
	
	
	\section{Introduction} \label{Intro} In recent times there has been a rapid development of the Internet of Things (IoT) that provides global sensing and computing capabilities to connect a broad range of devices to the Internet \cite{fl0}. These intelligent automated devices collect data from sensors and analyze such data at a centralized server. However, increased personalized requests from the user require such systems to analyze sensed data using a learning model. Learning the sensed data helps these systems to fulfil such requests, and enhances the quality of connected living. Traditional learning approaches follow a centralized architecture that allows the transmission of vast amounts of data from the device to a centralized server like a cloud. Such a transmission is vulnerable to malicious attacks, which may raise privacy concerns. In this context, federated learning (FL) introduces a decentralized learning architecture where user devices build their learning models using their onboard data and send it to the centralized server for preparing a globally learned model \cite{fl1}. Such a to-and-fro transmission of the learning model in FL eliminates the privacy breach of user data. On the other hand, the user devices update themselves from this globally learned model to cater to users' personalized needs. In this paper, we will extend more upon FL-as-a-service for multi-applications in IoT.
	
	In FL these globally learned models are shared within the connected devices to the cloud server, making it applicable for only one application. On the contrary, FLaaS is a system that enables different 3rd party applications to collaborate and build joint FL models to create high-level and extensible APIs for effectively meeting users’ personalized needs. To understand the user’s requirements, this collaborative training of FL models uses APIs on the user’s device to accurately find solutions. Despite the various advantages of FLaaS, there exist many challenges in its implementation \cite{fl2}.
	\begin{enumerate}
\item IoT devices grants applications and various services access to the user’s data. So far, FLaaS is unable to provide security and permission mechanisms across applications and services to share data and models among them. 
\item In FLaaS multiple applications and services can be involved in the collaborative modeling which leads to various privacy concerns of users’ data. A solution to this could be leveraging Differential Privacy (DP) noise at different stages of the FL system. Existing DP solutions in FLaaS affects the convergence rate of the FL model, thereby reducing the model’s utility.
\item In recent works, it has been proven that under a hierarchical design great convergence and accuracy can be achieved. So far, FLaaS does not have an optimal hierarchical architecture using FL methods across network layers.
\end{enumerate}

	\section{Related Work}\label{RW} This section examines numerous approaches and strategies proposed to improve the efficiency, performance, and fairness of the Federated Learning (FL) process in order to provide a full understanding of client selection. Client selection is a critical component in FL, as the choice of participating clients significantly impacts the overall model quality, training speed, and resource utilization. However, client selection faces several challenges, including handling network constraints, ensuring reliability, resource utilization and maintaining scalability in large-scale deployments. To address the network constraints issue, Albaseer et al. \cite{fl9} proposed a novel client selection algorithm to select clients with less latency issues and pushes for bandwidth reuse for clients that consume more time for model updation. Similarly, Zhang et al. \cite{fl10} and Xu et al. \cite{fl11} have proposed client selection algorithms that address the challenges of optimizing client selection and bandwidth allocation in FL networks, focusing on factors such as energy efficiency, learning performance, and long-term impact on model quality. 

    For ensuring reliability across the system, Wu et al. \cite{fl12} introduced HybridFL, a multi-layer federated learning protocol designed for Mobile Edge Computing (MEC) architectures, which addresses challenges such as optimizing efficiency, mitigating the impact of unreliable end devices, and improving FL training process outcomes significantly. Using a more unconventional method, Park et al. \cite{fl13} presented a novel model weight update method using Monte Carlo dropout to leverage local client reliability in Federated Learning-based indoor localization, improving performance compared to FedAvg and approaching centralized learning performance. Despite addressing the reliability challenge, the above works may face scalability issues when applied to large-scale systems.

    Scalability becomes an issue in client selection for FL in large-scale systems due to the increased complexity, resource constraints, communication overhead, and scalability challenges of algorithms. Lai et al. \cite{fl14} produced FedScale, an open-source federated learning benchmarking suite featuring realistic datasets, scalable runtime, and unified evaluation protocols, aiming to facilitate reproducible FL research and foster collaboration within the community. Whereas Li et al. \cite{fl15} introduced PyramidFL, which optimizes client selection through fine-grained utility profiling and achieves significant improvements in final model accuracy and training efficiency compared to state-of-the-art methods like Oort, as demonstrated on an open-source FL benchmark. Deng et al. \cite{fl16} addressed scalability and resource optimization challenge by proposing AUCTION, an Automated and Quality-aware Client Selection framework for efficient FL, designed to automatically evaluate client learning quality and select them with quality-awareness for FL tasks within resource constraints. AUCTION employs a neural network-based client selection policy encoded with reinforcement learning, adapting to dynamic client conditions and demonstrating efficiency, robustness, and scalability across real-world datasets and learning models.

    This paper provides a cohesive client selection algorithm to the reliability issues, resource constraints and scalability challenges in FLaaS for multi-application based systems.

\section{System Model}\label{SM} 
We consider an FL system consisting of client devices $CL$, access points $AP$, and a centralized cloud server. The network is modeled as a directed graph \[ G=(CL,L), \] where $CL$ denotes the set of client devices and $L$ denotes the set of communication links. Let $C_k \in CL$ represent the $k^{th}$ client device and $A_j \in AP$ represent the $j^{th}$ access point. The communication path between the client devices and the centralized cloud server is represented by $L_{ik} \in L$. The link between client device $C_k$ and access point $A_j$ is denoted by $l_{kj}\in L_{ik}$. We assume that each client device executes a single model computation task $t_k$ at a given time. The task is defined as \[ t_k := (w_k,s_k,q_k,d_k), \] where $w_k$ denotes the number of CPU cycles required to process one data sample, $s_k$ denotes the size of the local model update, $q_k$ denotes the number of local training iterations, and $d_k$ denotes the number of data samples available in the local dataset of client device $C_k$.
$$ $$

\noindent A. Delay Model
     \par We define decision parameter $y_{k}$ to represent whether the client device $C_{k}$ is selected $(y_{k}=1)$ or not $(y_{k}=0)$. Device selection in terms of latency can be decided by the local computation time and transmission delay. The time taken for local execution of task $t_{k}$ is given as \begin{equation} \tau_{k}^{loc}=\frac{w_{k}q_{k}d_{k}}{f_{k}} \label{eq:1} \end{equation} where $f_{k}$ represents the CPU frequency of the $k^{th}$ client device [4]. After local computation, all client devices upload their local FL models to the nearby access point via Frequency Division Multiple Access (FDMA). The maximum achievable rate between the $k^{th}$ client device and the $j^{th}$ access point is given by Shannon's equation as \[ r_{kj}=b_{k}\log_{2}\left(1+\frac{g_{kj}p_{k}}{N_{0}b_{k}}\right) \] where $b_{k}$ is the bandwidth allocated to the client device, $g_{kj}$ is the channel gain between the $k^{th}$ client device and the $j^{th}$ access point, $p_{k}$ is the transmission signal power of device $C_{k}$, and $N_{0}$ is the power spectral density of the Gaussian noise. We consider that device $C_{k}$ can access the network by associating with access point $A_{j}$ using an existing association policy $X$ such that $X(k)=j$. The time taken to transmit the model $s_{k}$ to the associated access point is given as \begin{equation} \tau_{k}^{tx}=\frac{s_{k}}{r_{k,X(k)}} \label{eq:2} \end{equation} We define decision parameter $x_{kj}^{k}$ for all $l_{kj}\in L_{ik}$ to denote whether link $l_{kj}$ is chosen for uploading task $t_{k}$. The link selection in terms of latency is determined by the propagation delay between the access point and the cloud server. Let $\tau_{ij}$ be the propagation delay associated with each link $l_{ij}$. The propagation delay incurred in offloading task $t_{k}$ is given as \begin{equation}  \tau_{k}^{prp}=\sum_{kj}\tau_{kj}x_{kj}\label{eq:3} \end{equation} Therefore, from (\ref{eq:1}), (\ref{eq:2}), and (\ref{eq:3}),we define the cost function for the delay model as \begin{equation} J_{\tau}(x,y):=\sum_{k}\left[\tau_{k}^{loc} +\tau_{k}^{tx} +\tau_{k}^{prp}\right]y_{k} \label{eq:delaycost}\end{equation}
$$ $$
\noindent B. Energy Consumption Model
          \par The device reliability in terms of energy consumption model can be decided by the local computation energy required to compute the local FL model is \begin{equation} \delta_{k}^{loc}=\kappa q_{k}w_{k}d_{k}f_{k}^{2} \label{eq:local_energy} \end{equation} where $\kappa$ is the effective switch capacitance of chip. Now the link reliability can be decided by the energy consumption of the IoT device while transmitting the data model $s_{k}$. The energy required for offloading task $t_{k}$ is given as \begin{equation} \delta_{k}^{tx}=p_k\tau_{k}^{tx} \label{eq:tx_energy} \end{equation}

Therefore, we define a cost function from (\ref{eq:local_energy}) and (\ref{eq:tx_energy}) for energy consumption model as \begin{equation} J_{\delta}(x,y):= \sum_k \left[ \delta_k^{loc} + \delta_k^{tx} \right]y_k \label{eq:energycost} \end{equation}
$$ $$
\noindent C. Link Model
         \par The link selection of client device $C_{k} \epsilon CL$ between access points $A_{j} \epsilon AP$ and cloud server can be measured by the traffic congestion, packet loss rate and noise factor. The packet loss rate of in a link $l_{ij} \epsilon L_{ik}$ is given as \begin{equation} \theta_{kj}^{plr} = \left[ \frac{N_{ij}^{tx}-N_{kj}^{rx}} {N_{kj}^{tx}} \right]100\% \label{eq:plr} \end{equation} where $N_{kj}^{tx}$ and $N_{kj}^{rx}$ is the total number of transmitted packets and received packets from access point to cloud server, respectively.

The network congestion of a link can be defined by following the NDN congestion control for multipath. This depends on the estimation of network congestion levels according to the changes of RTT (Round Trip Time) measured. The probability of congestion has a linear relationship with the real time value of RTT on an interval [$R_{min}$, $R_{max}$]. The probability is given as \begin{equation} \theta_{kj}^{prob}(t) = p_{min} + \Delta p_{max} \frac{R(t)-R_{min}(t)} {R_{max}(t)-R_{min}(t)} \label{eq:congestion} \end{equation}
where $\Delta p_{max} = p_{max} - p_{min}$, R(t) is the real-time value of RTT of currently sent Interest Package. When $RTT_{max} = RTT_{min} $ then \begin{equation} \theta_{kj}^{prob}(t)=p_{min}. \label{eq:minprob} \end{equation}

Therefore, from (\ref{eq:plr})--(\ref{eq:minprob}) we define a cost function for energy consumption model as \begin{equation} J_{\theta}(x,y):= \sum_k \left[ \theta_k^{plr} + \theta_{kj}^{prob} \right]y_k \label{eq:linkcost} \end{equation}
$$ $$

% \noindent D. Uniform Data Model
% \par To improve the global model’s convergence efficiency, we need to analyze each client’s contribution to the global model convergence in each round. According to information theory, the entropy of a random variable represents the average amount of information contained in the variable’s possible outcomes. Hence, we assume that the element in the gradient vector follows a certain random distribution and we calculate its entropy as an indicator of the amount of the convergence information contained in the local model parameters \cite{fl3}. Hence, when the server receives the locally updated models from the devices, it measures the entropy of each device k by utilizing the public data as
% ${\mu}(k) = \dfrac{1}{N_{p}}  \sum_{n=1}^{N_{p}} {\mu}_{x_{p, n}} (k)$ where $ {\mu}_{x_{p, n}} (k)$ is the shannon entropy of the model of the k-th device on the sample $x_{p, n}$ written as ${\mu}_{x_{p, n}} = - \sum_{q=1}^{Q} P_{x_{p, n}}^{(q)} log P_{x_{p, n}}^{(q)} (k)$. Here, Q is the number of classes of the dataset and $P_{x_{p, n}}^{(q)} (k)$ is the probability of prediction for the q-th class on a sample $x_{p, n}$ using the model of the k-th device \cite{fl4}. 

\section{Solution Approach}\label{SA}
 \noindent A. Problem Formulation
 \par To jointly design the client selection and the FL algorithm, we will now formulate an objective function for maximizing efficiency. We will take the minimization approach while factoring the entropy along with the client selection parameters.
Using the delay, energy, and link reliability cost functions defined in (\ref{eq:delaycost}), (\ref{eq:energycost}) and (\ref{eq:linkcost}) respectively, we formulate the client selection problem as:

\begin{equation} \begin{aligned} \underset{b,f,p,T,E}{\min} \quad & {\alpha_{1}}J_{\tau}(x,y) +{\alpha_{2}}J_{\delta}(x,y) +{\alpha_{3}}J_{\theta}(x,y) \\ \textrm{s.t.} \quad & x_{kj},y_k \in \{0,1\}, \qquad \forall k \in K,\\ & 0 \leq f_k \leq f_k^{\max},\\ & 0 \leq p_k \leq p_k^{\max},\\ & \sum_{k=1}^{K} b_k = B,\\ & 0 \leq J_{\tau}(x,y) \leq T_k^{\max},\\ & 0 \leq J_{\delta}(x,y) \leq E_k^{\max}. \end{aligned} \label{eq:problem_formulation} \end{equation}
  
 \par
      \emph{Theorem 1: }Our minimized FL client selection problem is NP-Hard.

 \par
      \emph{Proof: } Our optimization problem is a multiple integer noninteger linear programming (MINLP)\cite{fl5}. The objective is to maximize efficiency by minimizing the formula and taking the inverse. The 3 constraints in our formula are B, $f_{k}^{max}$, $p_{k}^{max}$ which is the total bandwidth, maximum computation capacity and maximum transmission power, respectively. Since, our minimization problem is a MINLP, hence it is NP-Hard\cite{fl6}. NP-hard problems cannot be solved using any heuristic algorithm. We can reduce our minimized problem into a NP problem. If our problem is NP-hard and NP, then it is an NP-complete problem, which can be solved using any heuristic algorithm. The following steps are given to reduce our problem to NP-complete problem:
      \begin{enumerate}
          \item The link optimization problem is first solved assuming an optimum path from client node to cloud.
          \item The device optimization problem is solved by assuming the optimum device from client node to cloud.        
      \end{enumerate}
$$ $$
\noindent B. Link Optimization
    \par
        For solving this problem we assume that the link path from client node to cloud has the shortest path, has no packet loss and suffers no latency during transmission and propagation. Taking this assumption into account, we take $ b_{k} $ and $ T_{k} $
        as constants so that makes the Link Model $J_{\theta}(x, y)$ and the Delay Model $J_{\tau}(x, y)$ a constant value as well.

        \begin{equation} \min_{f,p,E} \left[ \alpha_{2}J_{\delta}(x,y) + \alpha_{4}\mu(k) \right] + C, \label{eq:link_opt_obj} \end{equation}
        \newline s.t. C = ${\alpha_{1}}J_{\tau}(x, y) + {\alpha_{3}} J_{\theta}(x, y) $
        \newline $x_{kj}, y_{k} \epsilon {0,1} \hspace{1cm} \forall k \epsilon K$
        \newline $ 0 \leq f_{k} \leq f_{k}^{max}$
        \newline $ 0 \leq p_{k} \leq p_{k}^{max}$ 
        \newline $ 0 \leq J_{\delta}(x,y) \leq E_{k}^{max}$
$$ $$

\noindent C. Device Optimization
    \par
        For solving this problem we assume that the chosen client device consumes minimum energy and lowest entropy with the given data set. Taking this assumption into account, we take $ p_{k}, f_{k} $ and $ E_{k} $ as constants so that makes the Energy Model $J_{\delta}(x, y)$ and the Uniform Data Model $J_{\mu}(x, y)$ a constant value as well.

        \begin{equation} \min_{b,T} \left[ \alpha_{1}J_{\tau}(x,y) + \alpha_{3}J_{\theta}(x,y) \right] + C , \label{eq:device_opt_obj} \end{equation}
        \newline s.t. C = ${\alpha_{2}}J_{\delta}(x, y) + {\alpha_{4}} \mu(k) $
        \newline $x_{kj}, y_{k} \epsilon {0,1} \hspace{1cm} \forall k \epsilon K$
        \newline  $ \sum_{k=1}^{K} b_{k} = B $
        \newline $ 0 \leq J_{\tau}(x,y) \leq T_{k}^{max}$
        
        A greedy decentralized approach will be followed wherein for each task $t_k$ every client device $C_k$ $\forall k \epsilon K $ broadcasts their delay, energy and link cost function and data entropy to other client devices. Now one client device $C_0$ compares its cost function values with other client devices and stores its acceptance (1) or rejection (0) in a array called list. For each task, depending upon the number of 1s or 0s in the list, $y_k$ will be set and returned to the base station. Thereby giving us a clear indication whether a client is accepted or rejected for each task $t_k$. This approach is chosen as it significantly reduces the search operation to $K$ where K is total number of client devices.
        
        \begin{algorithm}
        	\SetAlgoLined
        	\KwIn{Data(k) = ($J_{\tau}^{k}(x,y), J_{\delta}^{k}(x,y), J_{\theta}^{k}(x,y), {\mu}(k)$), client device ($C_k$), Task ($t_k$) }
        	\KwOut{$y_k$}
        	    \For{task $t_k$ $\forall k \epsilon K$}{
                      Broadcast message with Data (k) to all $C_k$\;
                         \For{ $C_k$  $ 1 \leq k \leq K$}{
                            \uIf {$J_{\tau}^{0} < J_{\tau}^{k}$ \&\& $J_{\delta}^{0} < J_{\delta}^{k}$ \&\& $J_{\theta}^{0} < J_{\theta}^{k}$ \&\& ${\mu}(0) < {\mu}(k)$}{
                               set list[k] $\longleftarrow$ 1\; 
                            }
                            \Else{
                               set list[k] $\longleftarrow$ 0\;
                            }
                        }
                        Count 0s and 1s in the array list[]\;
                        Let a be number of 0s and b be number of 1s\;
                        \uIf {$a < b$}{
                               set $y_k$ $\longleftarrow$ 1\; 
                            }
                            \Else{
                               set  $y_k$ $\longleftarrow$ 0\;
                            }                  
                    }
          	    
                \Return $y_k$
        	\caption{\textbf{Greedy Decentralized Approach}}
        	\label{algo:aalgo}
        \end{algorithm}
        
	\section{Theoretical Analysis} \label{key}
 \par \textbf{Lemma 1. } \emph{The delay cost function $J_{\tau}^{k}(x,y)$, $\forall k \epsilon K$ will not decrease with the usage of buffer nodes}
 \par \emph{Proof: } Assuming there are 2 delay expressions $J_{\tau}^{k}(x,y) [with] $, $\forall k \epsilon K $ using buffer nodes and $J_{\tau}^{k}(x,y) [without] $, $\forall k \epsilon K $ without using buffer nodes and a timeout system. Let $J_{\tau}^{k}(x,y) [with] $, $\forall k \epsilon K $ be $D^{k}[with]$ and $J_{\tau}^{k}(x,y) [without] $, $\forall k \epsilon K $ be $D^{k}[without]$ for simplicity. Let us assume that from a pool of 100 nodes, we are to select 10 nodes using Greedy Decentralized Algorithm.

 $D^{k}[without] = \sum_iE(D_i^{k})$, where $E(D_i^{k})$ is the expected completion times on randomly selected nodes.
 $D^{k}[with] = \sum_iE'(D_i^{k}) + p * (Status Check Delay + Node Switching Delay)$, where $p$ is the probability of buffer nodes used.

 \textbf{Case 1:} Buffer nodes are not used\\
   $D^{k}[with] < D^{k}[without]$, in the proposed approach, the 10 nodes selected using greedy approach are chosen due to their low delays. The traditional approach uses central node to pick 10 best nodes which increases the time complexity. This contradicts the assumption.

 \textbf{Case 2:} Buffer nodes are used\\
   $D^{k}[with] = \sum_iE'(D_i^{k}) + p * (Status Check Delay + Node Switching Delay)$ \\
   $D^{k}[without] = \sum_iE(D_i^{k}) + (Remaining Time)$

   $D^{k}[with] < D^{k}[without]$

   Firstly, in the traditional approach without timeouts initiated by the central server to receive models from selected nodes, the system experiences an extended $Remaining Time$. This delay significantly surpasses the time needed for status checks and node switching. The introduction of buffers and timeouts in the proposed system effectively reduces this delay, enhancing overall system efficiency.

   Secondly, p, the probability of using buffers is small as we are using greedy where the best nodes are selected and very few are likely to fail. Therefore, $D^{k}[with]$ reduces.

   Hence in both the cases the $D^{k}[with] < D^{k}[without]$, thereby contradicting the assumption.

 \par \textbf{Lemma 2. } \emph{The energy cost function $J_{\delta}^{k}(x,y)$, $\forall k \epsilon K$ will increase.}
 \par \emph{Proof: }On taking the assumptions of Lemma 1, we can show that on increasing transmission signal power from $p_k$ to $p_k^{max}$, the offloading energy $\delta_{k}^{tx}$ increases as $p_k$ is directly proportional to it. Similarly on increasing CPU frequency from $f_k$ to $f_k^{max}$, the local computation energy $\delta_{k}^{loc}$ increases as $f_k$ is directly proportional to it.

 \par \textbf{Time Complexity: }The Greedy Decentralized algorithm has 2 loops and 2 if-else statements inside the for-loop. The overall time complexity using Big-Oh notation is O($t_k$ . $C_k$) $\forall k \epsilon K$. In short we can say that the time complexity is O($K^2$).
	
	\section{Performance Evaluation}\label{PE}
		\subsection{Experimental Design}\label{PE:ED}
  \par The experimental design of this study evaluates a model trained on the Human Activity Recognition Using Smartphones dataset, with Python used to implement and calculate key metrics, specifically delay and energy, based on predefined formulas. To simulate realistic network conditions, a link model was developed, using packet loss model parameters to assess the quality of node connections. With nodes randomly selected between 1 and 100, we analyzed their connectivity and performance over multiple rounds, yielding insights into delay and energy dynamics across varying network states.

\par To determine optimal node selection, three distinct cases were tested, each targeting a group of 10 nodes. The first case involved random selection based on delay and energy values alone. In the second, Q-learning was applied, allowing for iterative optimization by leveraging past performance data. The third approach employed a greedy algorithm, hypothesized to outperform the others by maximizing performance efficiency. By comparing these methods, the study underscores Greedy’s potential in achieving better delay and energy outcomes, affirming its utility in optimizing node selection in dynamic environments.
            \subsection{Performance Matrices }\label{PE:PM}
\par In this paper, we focus on key performance metrics that are critical for evaluating the effectiveness of our proposed Federated Learning (FL) approach: delay, energy consumption, and link status. Delay measures the time taken for data to be transmitted and processed, directly influencing the responsiveness of the system. Energy consumption indicates the resource efficiency of the nodes, essential for sustainable operation in IoT environments. Link status assesses the reliability of the connection between nodes and the central server, reflecting the stability of communication channels. To analyze these metrics, we conducted 40 experiments across 100 nodes, calculating the average values for each metric per experiment. This allows us to visualize the performance trends and draw meaningful comparisons across different node selection strategies in our study.
\begin{figure}[h] % 'h' places the figure here
    \centering
    \includegraphics[width=0.5\textwidth]{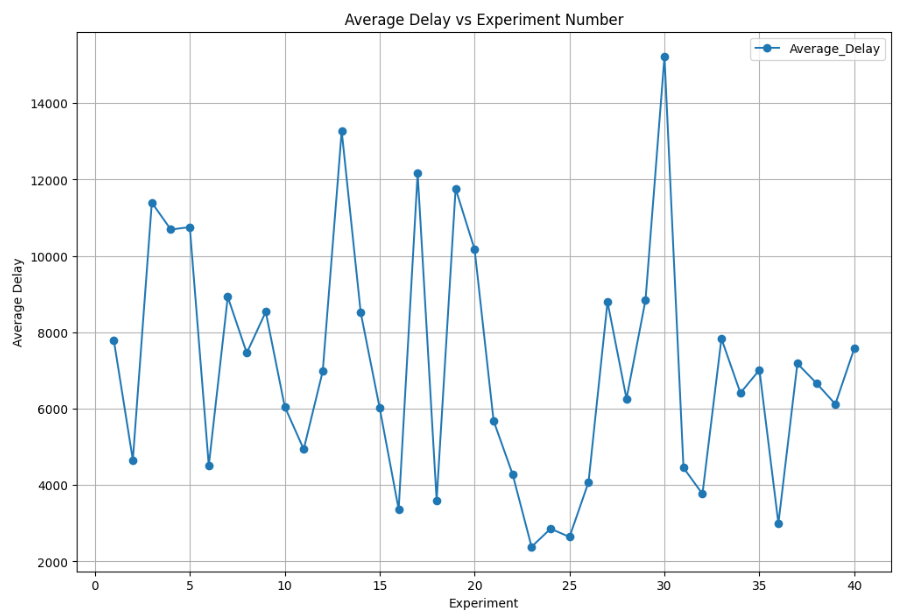} 
    \caption{Average Nodal Delay}
    \label{fig:delay}
\end{figure}

\par The graph in Fig \ref{fig:delay} displays the variation in average delay across 40 experiments, with each experiment measuring the time required to transmit data across nodes in a network. The delays fluctuate significantly, ranging from around 3,000 to over 14,000 units, indicating variability in network performance under different experimental conditions. Peaks observed in experiments 18, 23, and 30 suggest periods where transmission delays were notably higher, potentially due to network congestion, reduced link quality, or high energy consumption. Meanwhile, dips between experiments reflect more efficient data transfer scenarios, possibly due to favorable link conditions or optimized node performance. This fluctuation highlights the importance of monitoring delay as a key metric in evaluating network stability and performance.

\begin{figure}[h] % 'h' places the figure here
    \centering
    \includegraphics[width=0.5\textwidth]{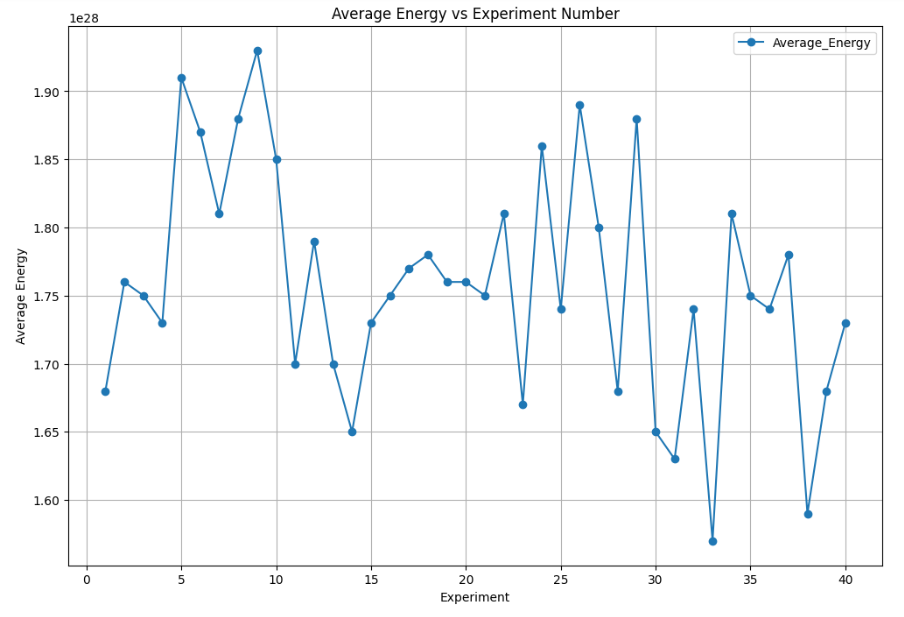} 
    \caption{Average Nodal Energy}
    \label{fig:nodal}
\end{figure}

\par The graph in Fig \ref{fig:nodal} illustrates the average energy consumption across 40 experiments, with values fluctuating between 1.60e+28 and 1.90e+28 units. There is noticeable variability in energy usage, particularly in the initial experiments, where peaks occur around experiments 5 and 8. This may indicate higher processing demands or inefficient energy use during these instances. The values stabilize slightly around 1.80e+28 in the middle range of experiments (15 to 30), followed by occasional dips, suggesting periods of more energy-efficient operation. Such fluctuations in energy metrics highlight the network’s inconsistency in power consumption, likely influenced by varying node conditions, link quality, or load distribution. This metric is essential for understanding the energy efficiency and sustainability of the network under different conditions.
  
		\subsection{Benchmark Schemes}\label{PE:BS}
\par In this paper, we utilize benchmark schemes to assess the performance of different node selection strategies based on key metrics such as delay, energy consumption, and link reliability. Specifically, we compare three selection approaches: a Greedy method, which prioritizes nodes expected to minimize delay and energy costs; a Q-Learning approach, which leverages reinforcement learning to adaptively select nodes based on cumulative experience; and a Random selection strategy, which serves as a baseline for evaluating the advantages of more structured methods. These benchmark schemes allow us to highlight the strengths and limitations of each approach, providing insights into which strategy best supports efficient and reliable model processing in distributed IoT systems.
\begin{figure}[h] % 'h' places the figure here
    \centering
    \includegraphics[width=0.5\textwidth]{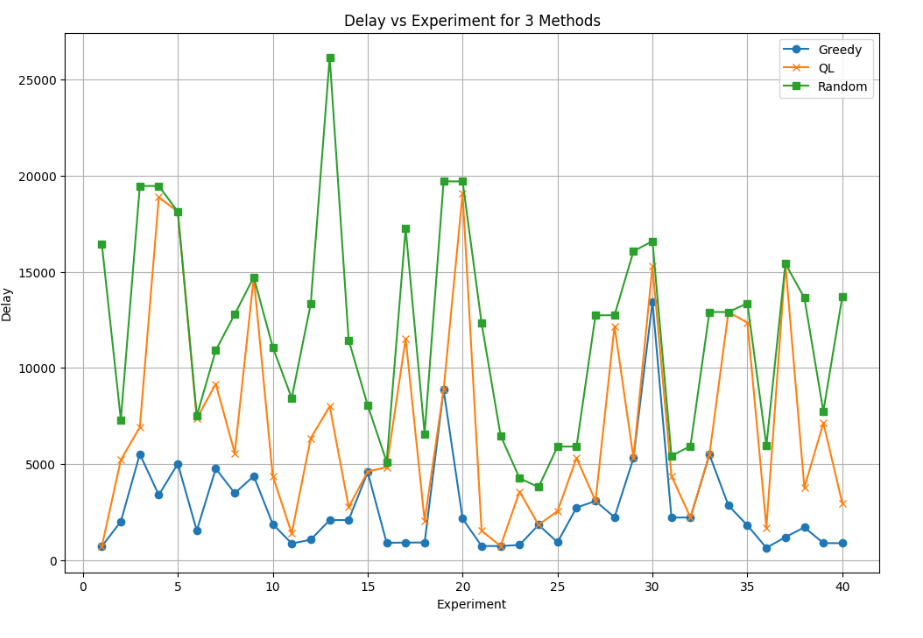} 
    \caption{Delay Comparison with Randomized approach, Q-Learning and Greedy approach}
    \label{fig:1}
\end{figure}

\par The graph compares the delay performance across 40 experiments using three methods: Greedy, Q-Learning (QL), and Random selection. The Greedy method consistently achieves the lowest delay across experiments, with values generally below 5,000, indicating a more efficient selection strategy. Q-Learning also performs reasonably well, though with higher delay values compared to Greedy, and shows variability with occasional spikes above 10,000, reflecting fluctuations in learning efficiency. The Random method exhibits the highest delays, frequently reaching or exceeding 20,000, indicating a lack of optimization. This comparison highlights the effectiveness of the Greedy method in minimizing delay, followed by Q-Learning, while Random selection results in significantly higher delays due to its lack of a targeted approach.

\begin{figure}[h] % 'h' places the figure here
    \centering
    \includegraphics[width=0.5\textwidth]{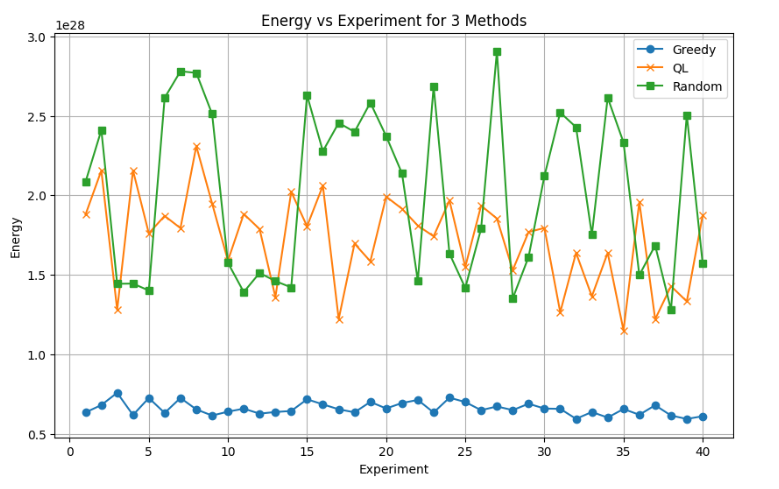} 
    \caption{Energy Comparison with Randomized approach, Q-Learning and Greedy approach}
    \label{fig:1}
\end{figure}

\par The graph presents a comparison of energy consumption across 40 experiments for three methods: Greedy, Q-Learning (QL), and Random. The Greedy method consistently consumes the least energy, with values slightly above 0.5e+28 and minimal fluctuation, indicating stable performance. The QL method shows moderately higher energy consumption, fluctuating between approximately 1.3e+28 and 2.2e+28, suggesting some variability. The Random method exhibits the highest variability and overall energy usage, with values ranging significantly between 1.5e+28 and 3e+28. This analysis highlights the Greedy method as the most energy-efficient, while the Random method is the least efficient and consistent in energy consumption across the experiments.

\subsection{Results}\label{PE:RD}
\par In the final results and discussion, we analyze the performance of the three methods: Greedy, Q-Learning (QL) and Random based on key metrics such as delay, energy consumption and link status across 40 experiments. The primary goal is to assess which method offers the most effective balance between efficiency and consistency, especially in the context of selecting nodes for model processing in a federated learning setup.

\par The Greedy method consistently achieves the lowest delay across the experiments, demonstrating a steady performance advantage over both QL and Random. QL performs relatively well but shows higher fluctuations than Greedy, indicating that while it is effective, it lacks the predictability of the Greedy approach. The Random method has the highest and most variable delay, which suggests that it is not suitable for applications where timely processing is crucial. These results highlight that Greedy provides the most reliable and lowest delay, making it ideal for latency-sensitive applications.

\par In terms of energy efficiency, the Greedy method also stands out as the best performer, maintaining a consistently low energy consumption with minimal fluctuations. QL consumes slightly more energy and shows moderate variability, reflecting its adaptive yet somewhat inconsistent nature. The Random method, however, has the highest and most inconsistent energy usage, which can be a drawback for energy-constrained environments. The results suggest that the Greedy method is most suitable for scenarios where energy conservation is essential, while QL could be considered if slight trade-offs in energy efficiency are acceptable.

\par Overall, the results clearly indicate that the Greedy method consistently outperforms QL and Random across all evaluated metrics, making it the optimal choice for federated learning in IoT or edge environments where delay, energy efficiency, and link quality are critical. QL, while competitive, shows higher variability and may require further tuning to achieve comparable stability. The Random method, due to its unpredictable performance, is the least favorable choice.
		
	\section{Conclusion}\label{Con}
\par In conclusion, this study emphasizes the transformative potential of Federated Learning (FL) as a service (FLaaS) in the realm of the Internet of Things (IoT). By exploring node selection strategies based on parameters such as delay, energy consumption, and link status, we have demonstrated the effectiveness of a greedy approach in optimizing model processing and enhancing resource utilization. However, to further improve the performance and applicability of FL in dynamic IoT environments, future research should consider integrating additional parameters, such as data heterogeneity which accounts for the variability in data distributions across devices. Addressing data heterogeneity can lead to more robust models capable of adapting to diverse conditions and ensuring equitable contributions from all nodes. Additionally, exploring advanced techniques like reinforcement learning and hybrid algorithms may yield further improvements in model accuracy and convergence rates. Ultimately, by continuing to refine node selection strategies and incorporating a broader range of influencing factors, we can enhance the efficacy of FL in delivering personalized and intelligent services within Society 5.0.

	\bibliographystyle{IEEEtran}
	\bibliography{Reference}

\end{document}